\documentclass[10pt,conference]{IEEEtran}
\IEEEoverridecommandlockouts

\usepackage[letterpaper, top=0.7in, bottom=1.0in, left=0.625in, right=0.625in]{geometry}
\usepackage[utf8]{inputenc} 
\usepackage[T1]{fontenc}    
\usepackage{times}          

\usepackage{amsmath,amssymb,mathtools,bm}

\usepackage{tabularx,array,booktabs,threeparttable}

\usepackage[caption=false,font=footnotesize]{subfig}
\usepackage{placeins}
\usepackage{graphicx}
\usepackage{subcaption}
\usepackage{subcaption}  
\usepackage{tikz}

\usepackage[font=it]{caption}

\usepackage{siunitx}
\usepackage{xcolor}
\usepackage{hyperref}

\usepackage[numbers]{natbib}
\usepackage{float}

\def\BibTeX{{\rm B\kern-.05em{\sc i\kern-.025em b}\kern-.08em
    T\kern-.1667em\lower.7ex\hbox{E}\kern-.125emX}}

\begin{document}

\title{RF Intelligence for Health: Classification of SmartBAN Signals in overcrowded ISM band}

\author{
\IEEEauthorblockN{Nicola Gallucci\IEEEauthorrefmark{1}$^1$, Giacomo Aragnetti\IEEEauthorrefmark{1}$^1$, Matteo Malagrinò\IEEEauthorrefmark{1}$^1$,
\\ Francesco Linsalata\IEEEauthorrefmark{1}$^2$, 
\vspace{-0.4cm}Maurizio Magarini\IEEEauthorrefmark{1}$^2$, 
Lorenzo Mucchi\IEEEauthorrefmark{2} } \\
\IEEEauthorblockA{\IEEEauthorrefmark{1}DEIB, Politecnico di Milano, Milan, Italy, 
$^1$\textit{name.surname@mail.polimi.it}, $^2$\textit{name.surname@polimi.it}} \\
\vspace{-0.45cm}\IEEEauthorblockA{\IEEEauthorrefmark{2}\textit{Dept. of Information Engineering, Università di Firenze, Florence, Italy, lorenzo.mucchi@unifi.it}}
}

\maketitle
\begin{abstract}
Accurate classification of Radio-Frequency (RF) signals is essential for reliable wearable health-monitoring systems, providing awareness of the interference conditions in which medical protocols operate. In the overcrowded 2.4\,GHz ISM band, however, identifying low-power transmissions from medical sensors is challenging due to strong co-channel interference and substantial power asymmetry with coexisting technologies. This work introduces the first open source framework for automatic recognition of SmartBAN signals in Body Area Networks (BANs). The framework combines a synthetic dataset of simulated signals with real RF acquisitions obtained through Software-Defined Radios (SDRs), enabling both controlled and realistic evaluation. Deep convolutional neural networks based on ResNet encoders and U-Net decoders with attention mechanisms are trained and assessed across diverse propagation conditions. The proposed approach achieves over 90\% accuracy on synthetic datasets and demonstrates consistent performance on real over-the-air spectrograms. By enabling reliable SmartBAN signal recognition in dense spectral environments, this framework supports interference-aware coexistence strategies and improves the dependability of wearable healthcare systems.
\end{abstract}




\section{Introduction}

Body Area Networks (BANs) are becoming essential in modern healthcare, enabling continuous and non-invasive monitoring of vital signs through wearable medical devices \cite{ETSIMedicalIoT}. These technologies are increasingly integrated into daily life, with smartwatches and fitness trackers as familiar examples \cite{ETSIsmartban, ETSI-TC}.

\begin{figure}[!t]
\centering
{\includegraphics[width=0.8\columnwidth]{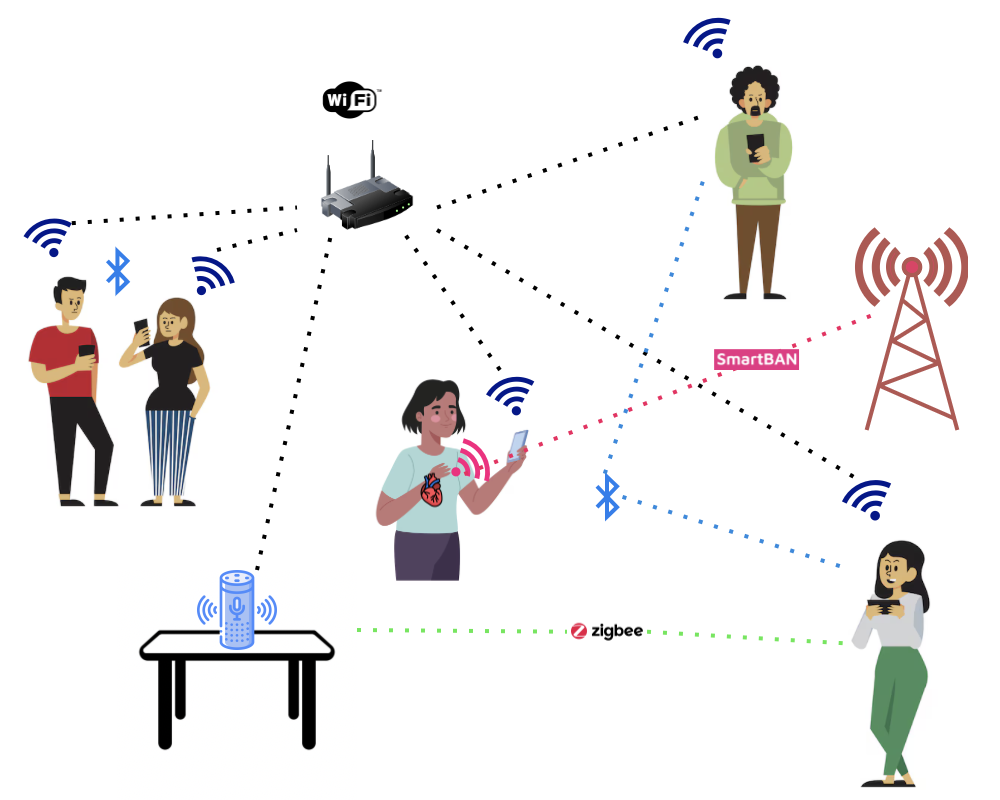}
\caption{An example of a crowded RF signals environment with different co-existing technologies at ISM band.}
\label{fig:scenario}
}
\end{figure}

To ensure interoperability and efficient spectrum usage, the ETSI Smart Body Area Network (SmartBAN) technical committee was established in 2013 to define standards for ultra-low-power physical (PHY) and medium access control (MAC) layers, as well as lightweight data formats \cite{viittala2017etsi, ETSI-TC}. The latest specifications include an ultra-low-power PHY (TS 103 326 \cite{etsi103326}), a low-complexity MAC (TS 103 325 \cite{etsi103325}), and interoperable data representation and management formats. SmartBAN typically adopts a star network topology, where a central coordinator manages control and network operations while acting as a gateway to external systems. Intelligence is further introduced through semantic data models and coexistence management strategies, such as cognitive channel sensing and dynamic frequency selection \cite{viittala2017etsi}.  

Despite these advances, SmartBAN transmissions operate in the crowded 2.4–2.48\,GHz ISM band, where they often coexist with higher-power technologies such as WLAN, ZigBee, and Bluetooth, as depicted in Fig.~\ref{fig:scenario}. The extremely low power of SmartBAN medical signals makes them particularly vulnerable to interference, threatening the reliable delivery of critical health data \cite{CoexistenceBAN}.

AI-based signal detection methods have garnered significant attention for wireless classification in dense ISM environments. Previous studies have applied machine learning techniques such as support vector machines (SVMs) and convolutional neural networks (CNNs) to recognize Wi-Fi, ZigBee, and Bluetooth transmissions, even in overlapping conditions. For instance, Rashidpour and Bahramgiri employed SVM and K-Nearest Neighbors algorithms to classify Wi-Fi and Bluetooth protocols, achieving classification accuracies of 97.83\% and 98.12\%, respectively \cite{rashidpour2024mac}. Similarly, He et al. utilized deep learning techniques to identify ISM band signals, demonstrating the efficacy of deep learning in this domain \cite{he2020ism}.
However, these approaches primarily target relatively strong signals with distinct spectral features. In contrast, SmartBAN medical waveforms present a more challenging scenario due to their extremely low power density and subtle spectral footprint. 

 Recent advancements have sought to address these challenges. For example, Zhang et al. \cite{zhang2021_shared_spectrum_classification} developed deep neural networks to detect coexisting signal types based on I/Q samples, achieving competitive accuracies even in the presence of multiple signal types. Additionally, Xu et al. proposed a composite time-frequency analysis method combined with a Siamese neural network for interference classification in frequency hopping systems, enhancing classification accuracy in complex environments \cite{xu2022time}. 

Furthermore, Software-defined radios (SDRs) enable realistic, over-the-air evaluation of classification networks by generating and capturing signals under real channel conditions, hardware impairments, and coexisting interference.
 A notable SDR-validated study by Zhang et al. \cite{zhang2021_shared_spectrum_classification} transmits coexisting Wi-Fi, LTE, and 5G New Radio compliant waveforms with USRPs and classifies raw I/Q and signal features using a convolutional neural network architecture, reporting high accuracy in over-the-air tests. However, these evaluations focus on relatively strong wireless standards and favorable signal conditions; in contrast, our work targets extremely low-power SmartBAN medical waveforms in realistic ISM coexistence scenarios.

These studies underscore the potential of AI-based methods in classifying weak and overlapping signals in ISM bands and the notable use of SDRs for real testing. However, the classification of SmartBAN signals remains an open challenge due to their low power density and subtle spectral characteristics. Further research is needed to develop robust AI-based frameworks capable of accurately detecting and classifying such weak signals in dense ISM environments. 
In this context, distinguishing SmartBAN transmissions from stronger coexisting signals is a key enabler for interference-aware spectrum access. Accurate classification provides the foundation for adaptive channel selection and coexistence control, ensuring reliable data exchange in wearable medical applications operating under severe power asymmetry and dense spectral overlap.

\textbf{Paper Contributions} To address this issue, we propose \textit{Smart RF Signal Recognition}, an open source \cite{smartban_repo} deep learning framework for the detection and classification of SmartBAN medical waveforms. The main contributions of this work are as follows:  
\begin{itemize}
    \item Provide an analysis of the spectral behaviour of SmartBAN transmissions in the 2.4--2.48\,GHz ISM band, emphasizing the difficulties introduced by strong co-channel interference from WLAN, ZigBee, and Bluetooth.  
    \item Design multiple neural architectures that combine ResNet encoders with U-Net decoders enhanced by attention mechanisms, optimized for dynamic and heterogeneous channel conditions.  
    \item Develop a hybrid evaluation methodology that integrates synthetic spectrogram simulations with real-world RF captures, enabling comprehensive assessment in mixed-signal environments.  
    \item Demonstrate the classification performance and robustness of the proposed framework, even under adverse channel conditions.  
\end{itemize}

\textbf{Paper Organization} The rest of this paper is structured as follows: Section~II describes the generation of the dataset used for training and evaluating the proposed deep learning framework. Building on the system model introduced in the previous section, Section~III presents the design of a deep learning framework for the semantic segmentation of spectrogram images. Section~IV reports the numerical and experimental results obtained using both synthetic and realistic spectrograms. Finally, Section~V concludes the paper.

\section{Signal and Interference Model}
\label{sec:signal_model}

In this section, we describe the generation of the dataset used for training and evaluating the proposed deep learning framework. The dataset is designed to capture the coexistence of multiple wireless technologies within the congested 2.4--2.48\,GHz ISM band.

\subsection{Time-Frequency Signal and Channel Model}

Let $s_i(t)$ denote the baseband waveform of the $i$-th wireless technology, with $i \in \{\text{Wi-Fi, Bluetooth, ZigBee, SmartBAN}\}$. Each signal propagates through a multipath fading channel $h_i(t,\tau)$ and is affected by additive white Gaussian noise $n(t)$. The received signal is expressed as
\begin{equation}
r(t) = \sum_{i=1}^{N} \int s_i(t-\tau) h_i(t,\tau) d\tau + n(t),
\end{equation}
where $N$ is the number of simultaneously active signals, and $h_i(t,\tau)$ represents the time-varying channel impulse response.

The fading channel is modeled in a general form to encompass both Rayleigh and Rician scenarios:
\begin{equation}
h_i(t,\tau) = \sum_{k=1}^{L_i} a_{i,k}(t) \, \delta(\tau - \tau_{i,k}),
\end{equation}
where $L_i$ is the number of multipath components, $\tau_{i,k}$ are the path delays, and $a_{i,k}(t)$ are the complex path gains.  
The channel coefficients $a_{i,k}(t)$ can be generally expressed as
\begin{equation}
a_{i,k}(t) = \sqrt{\frac{K}{K+1}} \, a_\text{LOS} + \sqrt{\frac{1}{K+1}} \, a_\text{NLOS}(t),
\end{equation}
where $K \ge 0$ is the Rician factor, $a_\text{LOS}$ is the deterministic line-of-sight (LOS) component (computed according to a log-normal path loss model), and $a_\text{NLOS}(t) \sim \mathcal{CN}(0, \sigma^2)$ represents the random scattered multipath component.

Rayleigh fading corresponds to the special case $K = 0$, in which the LOS component is absent and the channel is purely scattered.

To emulate realistic propagation conditions, each transmitted signal is subjected to a randomly selected channel model—Rician and Rayleigh—so as to reproduce diverse and dynamic environments.  
For each received signal, path loss is applied according to the transmission type.  

For WLAN transmissions, the path loss follows the indoor model defined in the 3GPP TR 38.901 report \cite{3gpp_TR_38.901}:
\begin{equation}
PL_{\text{WLAN}}(d) = 32.4 + 17.3 \log_{10}(d) + 20 \log_{10}(f_c),
\end{equation}
where $d$ is the transmitter–receiver distance (in meters) and $f_c$ is the carrier frequency (in GHz).  

Conversely, for SmartBAN, Bluetooth, and ZigBee transmissions, a log-normal path loss model is adopted, where the path loss exponent $n$ is set to 2.5 for LOS conditions and 3.5 for NLOS ones.

To map the received signals to the spectrogram-based label masks, we define the short-time Fourier transform (STFT) of the received signal:
\begin{equation}
R(t,f) = \int r(\tau) w(t-\tau) e^{-j2\pi f \tau} d\tau,
\end{equation}
with $w(t)$ a suitable hamming window function. The corresponding power spectrogram is
\begin{equation}
S_r(t,f) = |R(t,f)|^2,
\end{equation}
which serves as input to the neural network.

Similarly, the STFT of each channel $h_i(t,\tau)$ is
\begin{equation}
H_i(t,f) = \int h_i(t,\tau) w(t-\tau) e^{-j2\pi f \tau} d\tau,
\end{equation}
allowing us to approximate the received spectrogram as
\begin{equation}
S_r(t,f) \approx \left| \sum_{i=1}^{N} S_i(t,f) \cdot H_i(t,f) + N(t,f) \right|^2,
\end{equation}
where $S_i(t,f)$ is the STFT of the transmitted waveform $s_i(t)$ and $N(t,f)$ is the noise in the time-frequency domain.

\subsection{Signal Types and Composite Spectrograms}

The dataset includes four signal classes: Wi-Fi (IEEE 802.11ax), Bluetooth (IEEE 802.15.1), ZigBee (IEEE 802.15.4), and SmartBAN. Each waveform is parameterized according to nominal symbol rate, time span, packet/slot duration, and idle intervals, as summarized in Table~\ref{tab:parametri_segnali}.

Composite signals are generated by randomly combining one to four transmissions, allowing overlap in both time and frequency. Each component experiences distinct channel conditions, and additive white Gaussian noise (AWGN) is introduced to the overall signal, capturing realistic interference patterns. The received composite signals are converted to spectrograms using the STFT, producing the input images for training.

All spectrograms and corresponding masks are stored for training and evaluation. As explained in the following, the testing dataset is further augmented with real-world RF captures to improve generalization and validate performance under realistic interference and propagation conditions.

\begin{table}[t!]
\centering
\resizebox{\columnwidth}{!}{%
\begin{threeparttable}
\begin{tabular}{lllll}
\toprule
\textit{Parameters} & \textit{Bluetooth} & \textit{ZigBee} & \textit{Wi-Fi} & \textit{SmartBAN} \\
\midrule
Sample Rate & 80 MHz  & 80 MHz & 80 MHz & 80 MHz \\
Symbol Rate & 1 MHz & 4 MHz & 20 MHz & 1 MHz \\
Time Span  & 20 ms & 20 ms & 20 ms & 20 ms \\
Duration & 625 $\mu$s** & 4.2565 ms* & 180 $\mu$s* & 1.25 ms** \\
Idle Time & 625 $\mu$s & 0--5 $\mu$s & 20 $\mu$s & - \\
\bottomrule
\end{tabular}
\begin{tablenotes}
\small
\item[*] Packet duration; \item[**] Slot duration (Bluetooth and SmartBAN may use 1, 3, or 5 slots)
\end{tablenotes}
\end{threeparttable}
} 
\caption{Signal parameters used for dataset creation.}
\label{tab:parametri_segnali}
\end{table}

\begin{figure*}[t]
    \centering
    \begin{minipage}[t]{0.19\textwidth}
        \centering
        \includegraphics[width=\linewidth]{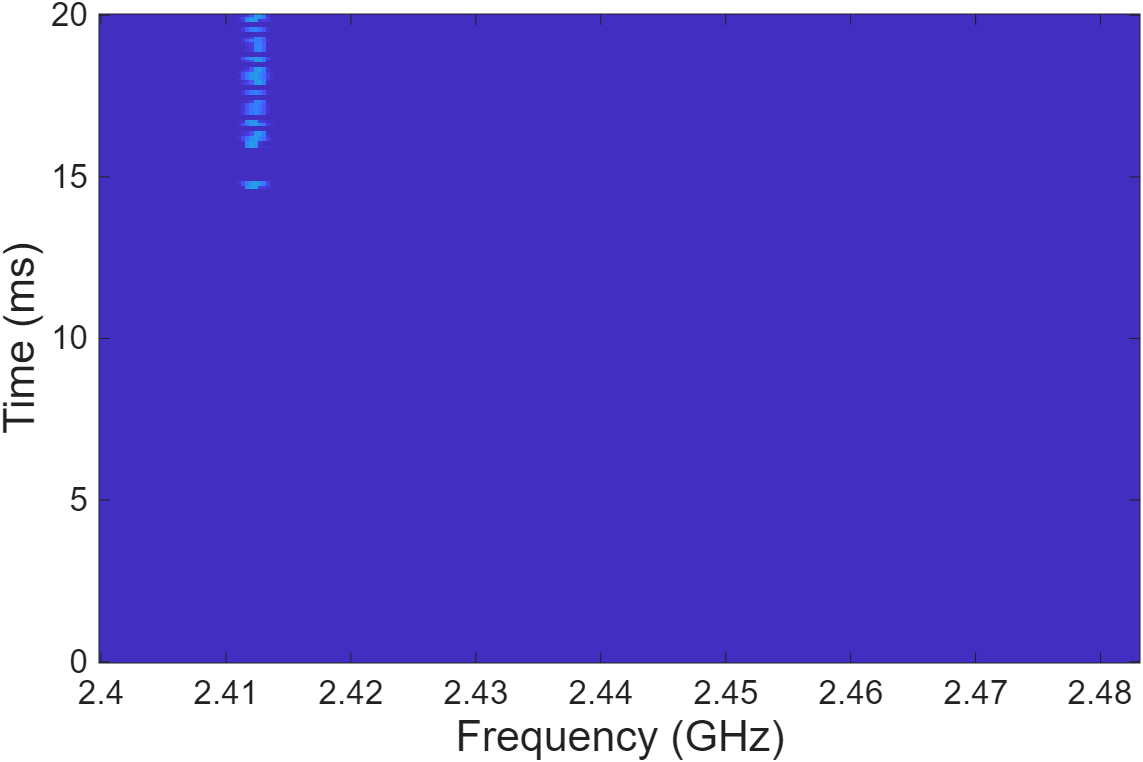}
        \caption*{(a) SmartBAN only}
    \end{minipage}\hfill
    \begin{minipage}[t]{0.19\textwidth}
        \centering
        \includegraphics[width=\linewidth]{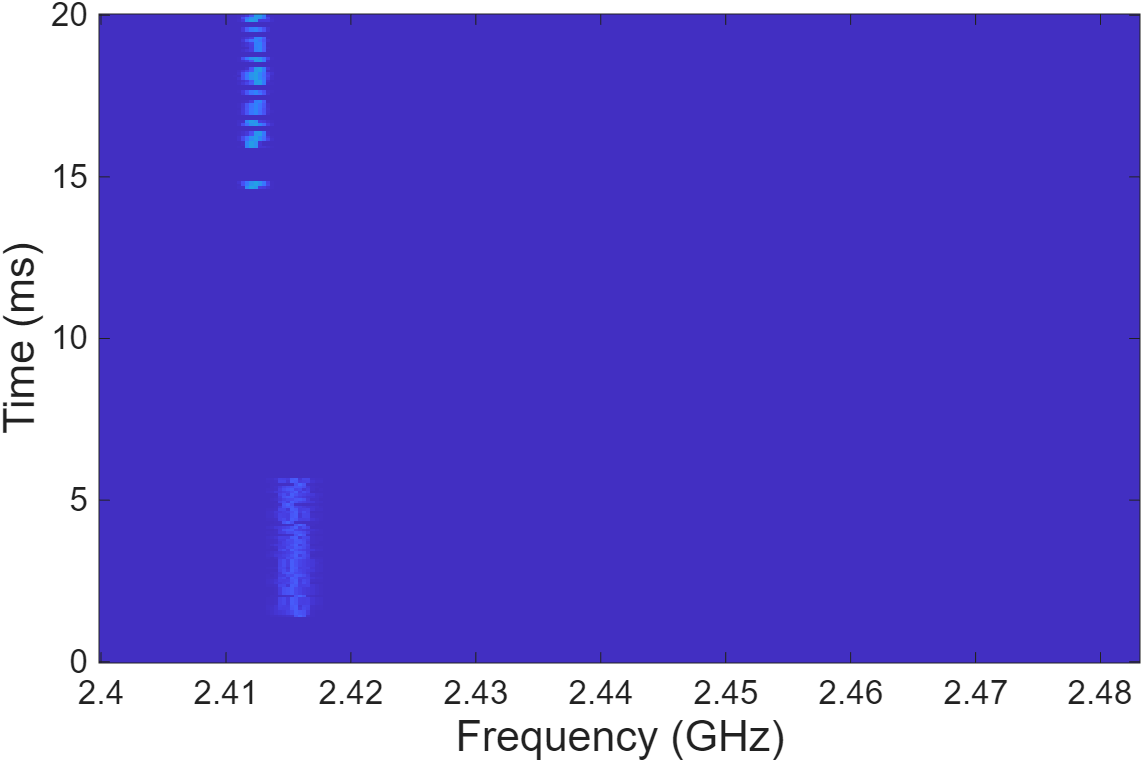}
        \caption*{(b) with 1 interferer}
    \end{minipage}\hfill
    \begin{minipage}[t]{0.19\textwidth}
        \centering
        \includegraphics[width=\linewidth]{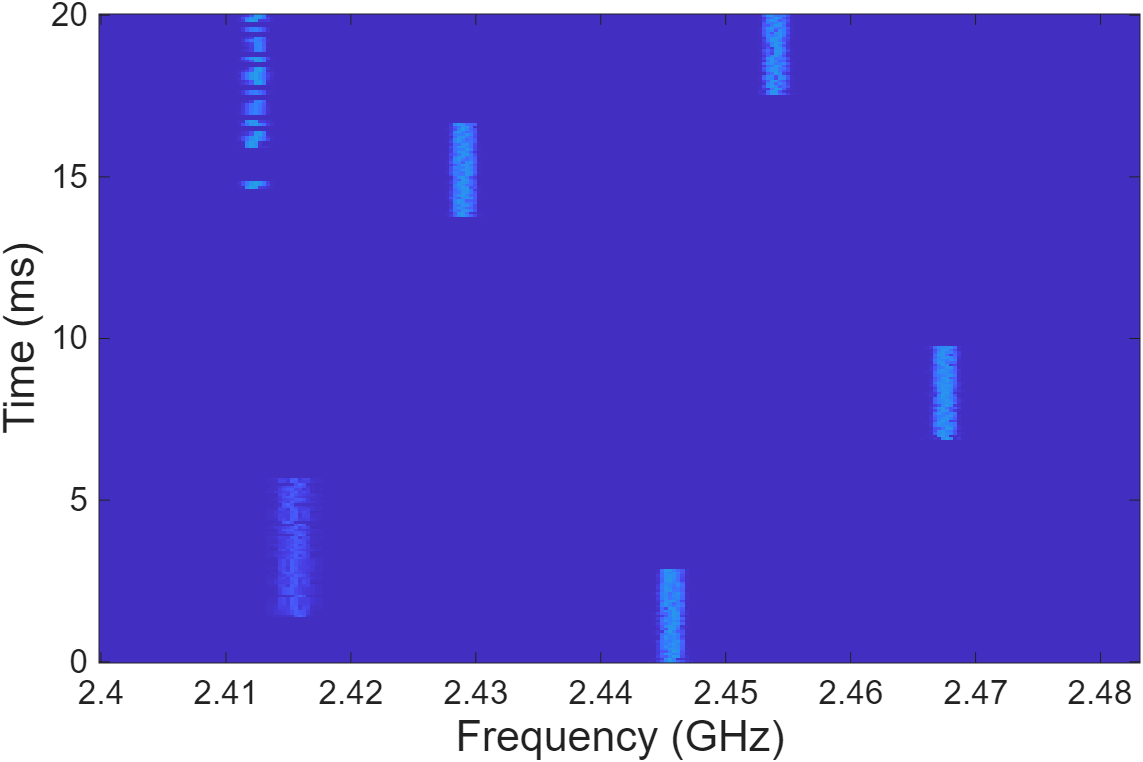}
        \caption*{(c) with 2 interferers}
    \end{minipage}\hfill
    \begin{minipage}[t]{0.19\textwidth}
        \centering
        \includegraphics[width=\linewidth]{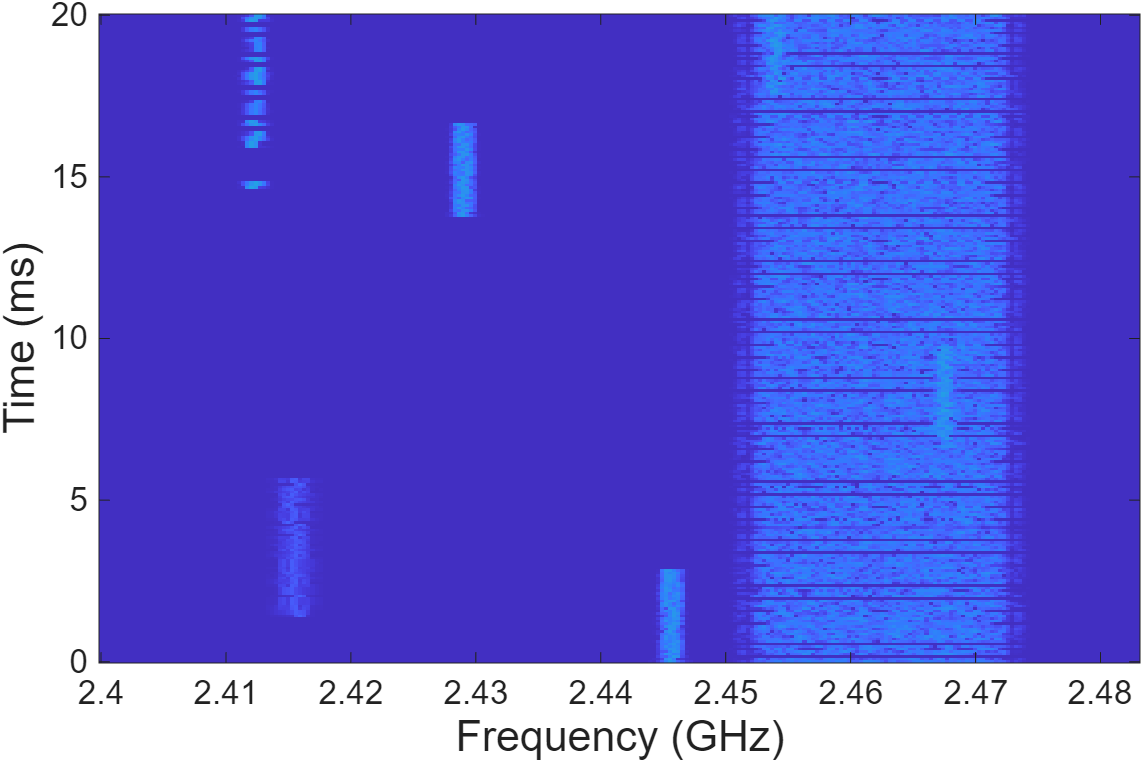}
        \caption*{(d) with 3 interferers}
    \end{minipage}\hfill
    \begin{minipage}[t]{0.21\textwidth}
        \centering
        \includegraphics[width=\linewidth]{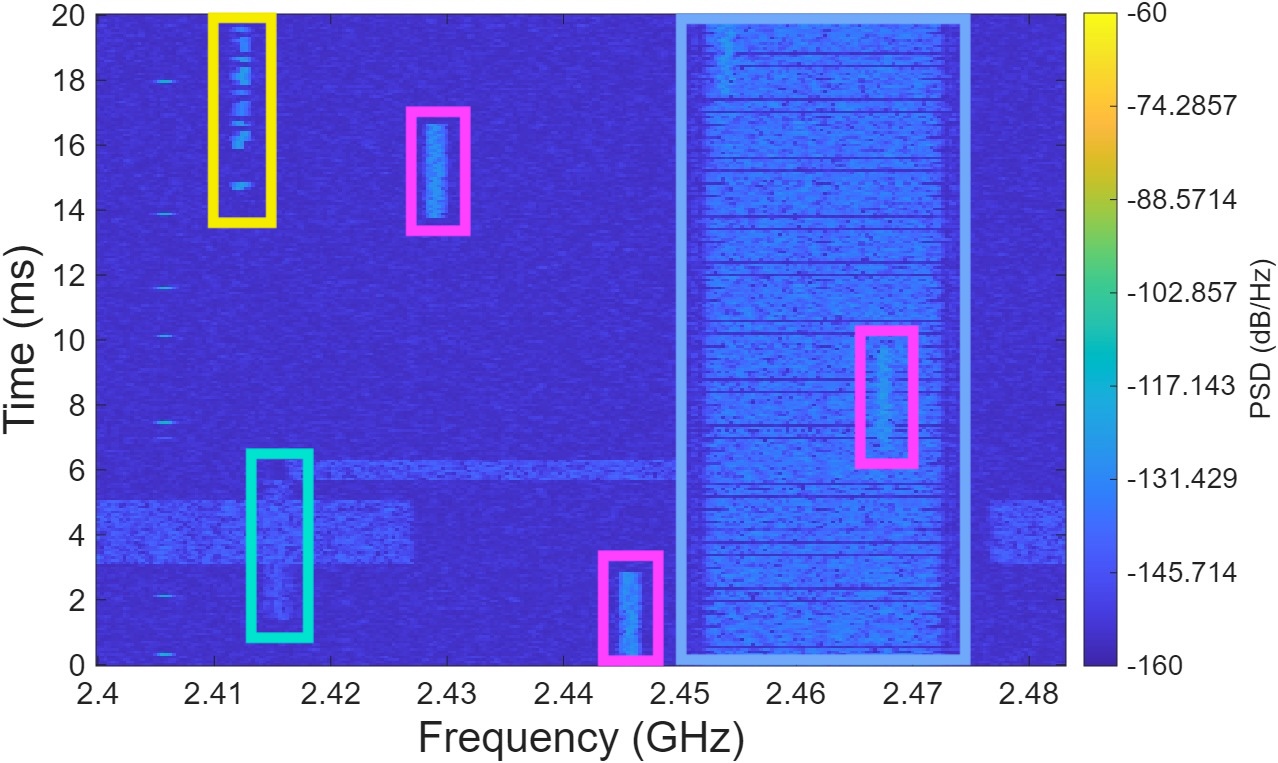}
        \caption*{(e) with 4 interferers \footnotemark}
    \end{minipage}
    \caption{Spectrograms of SmartBAN signal and increasing signal interference conditions. In (e) the yellow box stands for SmartBAN, water green box for ZigBee, purple box for Bluetooth, and light blue box for WLAN.}
    \label{fig:smartban_interference}
\end{figure*}

\section{Spectrogram-based Semantic Segmentation}

Building on the system model described in previous section, we design a deep learning framework for semantic segmentation of spectrogram images. Each time-frequency bin $(t,f)$ of the received spectrogram $S_r(t,f)$ is mapped to a label mask $M(t,f)$, indicating the dominant signal class.

\subsection{Model Design}

The framework supports encoder-decoder architectures.
The \textit{U-Net} with skip connections and attention gates is used as the decoder, whereas \textit{DeepLab v3+} with atrous spatial pyramid pooling (ASPP) or \textit{ResNet-18} or \textit{ResNet-50} as encoder, balancing depth and computational cost.


To enhance spatial focus, additive attention gates are applied on skip connections. Formally, with encoder features $F_e(t,f)$ and decoder features $F_d(t,f)$, the attention map $A(t,f)$ is computed as
\begin{equation}
A(t,f) = \sigma \big( W^T [F_e(t,f), F_d(t,f)] + b \big),
\end{equation}
where $\sigma$ is the sigmoid function and $W,b$ are learnable. The gated features are
\begin{equation}
\tilde{F}_e(t,f) = A(t,f) \odot F_e(t,f),
\end{equation}
with $\odot$ denoting element-wise multiplication. This mechanism suppresses irrelevant background activations, improving detection of overlapping and low-power signals such as SmartBAN. Furthermore, the inclusion of attention gates significantly improves the BF Score by enhancing the delineation of the signals. Indeed, the results on real-world captures using the same network architecture without attention gates show that the signal boundaries are indistinct and the resulting estimation is highly noisy.

The network is trained with a weighted pixel-wise cross-entropy loss to handle class imbalance:
\begin{equation}
\mathcal{L}_\text{CE} = - \sum_{t,f} w_{c(t,f)} \, M_{c(t,f)}(t,f) \log \hat{M}_{c(t,f)}(t,f),
\end{equation}
where $c(t,f) = M(t,f)$, $\hat{M}_{c(t,f)}(t,f)$ is the predicted probability, and $w_c$ is the inverse frequency of class $c$ in the training set.


%
\subsection{Training and Evaluation}
The dataset consists of 25,000 spectrograms, each representing a different wireless environment scenario. 
The data were generated by randomly distributing devices of various types within a circular area of radius \( 20\,\mathrm{m} \).  
A random number of devices,from one to four,  is selected for each scenario, and each device transmits a randomly chosen signal type among SmartBAN, WLAN, Bluetooth, and ZigBee.  
In the case of a single device, this is always set to transmit a SmartBAN signal to increase the occurrence of the target signal in the dataset. An example showing the spectrogram under increasing levels of interference, from one interferer up to all kinds of interferers, is shown in Figure~\ref{fig:smartban_interference}.


Spectrograms were generated from the input waveforms using the STFT with an FFT length of 4096, a Hann window of length 256 samples, and an overlap of 100 samples. Subsequently, the Power Spectral Density (PSD) values were converted to a logarithmic scale (dB), clipped to a dynamic range of $[-130, -50]$ dB, and finally min–max normalized to the unit interval $[0, 1]$.

Consequently, a \textit{ground truth matrix} of size \(256 \times 256\) is created for each spectrogram.  
Each pixel in the matrix corresponds to a specific signal class, encoded as an integer label: 
\(0\) for \textit{Unknown}, \(16\) for \textit{WLAN}, \(32\) for \textit{Bluetooth}, \(64\) for \textit{ZigBee}, and \(128\) for \textit{SmartBAN}.
In cases where multiple signals overlap in the time-frequency domain, a priority rule that highlights smaller signals is applied to assign a unique label to each pixel in the global ground truth. The priority rule is the following: \text{Bluetooth} > \text{SmartBAN} > \text{ZigBee} > \text{WLAN}.
This ensures unambiguous labeling even in overlapping regions prioritizing narrower-bandwidth signals.
The total spectrogram is then generated by combining all the signals, and the corresponding ground truth matrix is stored for subsequent processing.


The dataset is split into training (70\%), validation (20\%), and test (10\%) subsets, preserving the class distributions. Data augmentation includes additive noise and random overlapping of interfering signals to emulate congested ISM-band conditions. Training is performed using the Adam optimizer with stepwise decay of the learning rate. The network parameters $\theta$ are updated to minimize $\mathcal{L}(\theta)$ in the training set. All the training parameters are shown in Table \ref{tab:training_params}.

The evaluation of the test set compares the predicted segmentation maps \(\hat{M}(t,f)\) with the ground truth \(M(t,f)\) using several metrics. 
The \textit{Pixel Accuracy} measures the fraction of correctly classified time–frequency bins, while the \textit{Mean Intersection-over-Union (mIoU)} provides a balanced assessment of precision and recall. 
The \textit{Weighted F1-score} is employed to account for class imbalance, and \textit{class-wise confusion matrices} are analysed to identify the most common misclassification.

\begin{table}[!t]
\centering
\begin{tabular}{@{}ll@{}}
\textit{Parameter} & \textit{Value} \\
\midrule
Optimizer & Adam \\
Initial learning rate & $8 \times 10^{-4}$ \\
Learning rate decay & $0.1$ every 10 epochs \\
Total epochs & 25 \\
Mini-batch size & 32 \\
\end{tabular}
\vspace{2mm}
\caption{Training hyperparameters used in the segmentation pipeline.}
\label{tab:training_params}
\end{table}


\begin{table}[t!]
\centering
\small 
\begin{threeparttable}
\begin{tabularx}{\linewidth}{l 
  >{\raggedleft\arraybackslash}X 
  >{\raggedleft\arraybackslash}X}
\toprule
\textit{Network} & \textit{Prediction Mean (s)} & \textit{Prediction Variance} \\
\midrule
ResNet18      & 0.2846  & 0.0027 \\
ResNet50      & 1.760  & 0.0648 \\
DeepLab v3+   & 0.2779  & 0.0008 \\

\bottomrule
\end{tabularx}
\begin{tablenotes}
\vspace{1mm}
\footnotesize
\item[*] Simulations conducted on a MacBook Pro with Apple M1 chip.
\end{tablenotes}
\end{threeparttable}
\vspace{3mm}
\caption{Average prediction time and variance for each network}
\label{tab:prediction_time}
\end{table}

\begin{figure}[!t]
\centering
{\includegraphics[width=1\columnwidth]{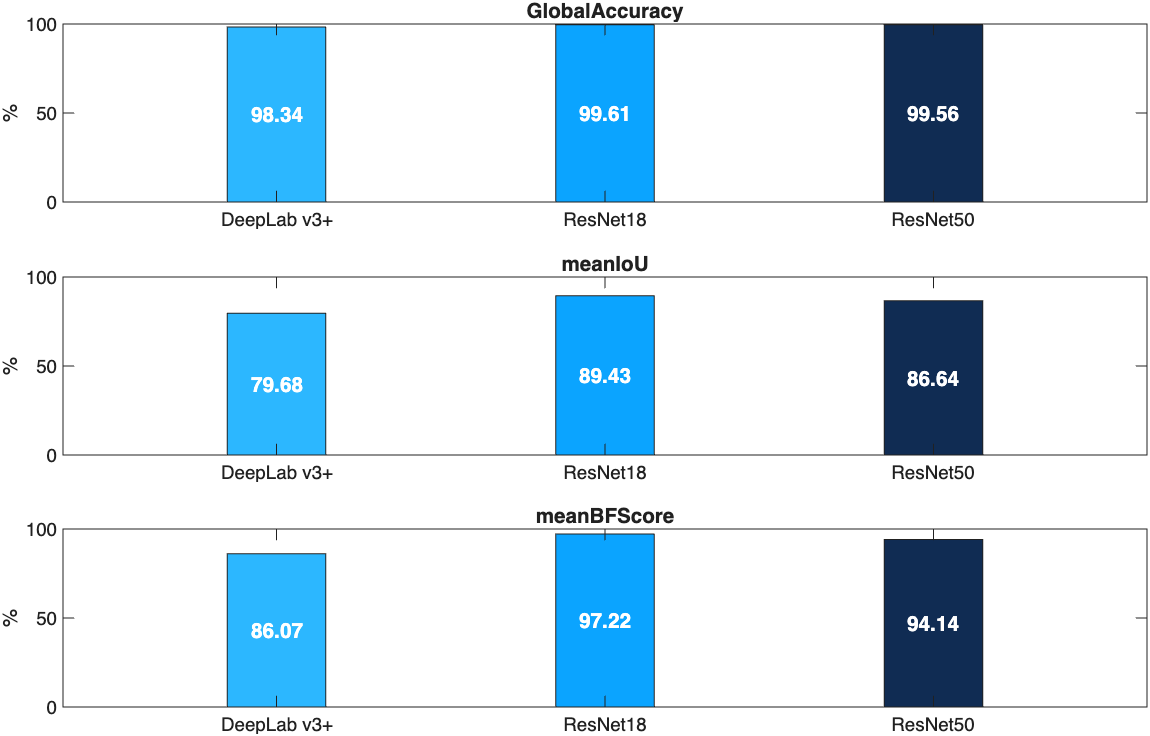}
\caption{Comparative analysis of performance metrics for different network architectures.}
\label{fig:comparativeAnalysis}
}
\end{figure}
\begin{figure}[!t]
\centering
{\includegraphics[width=0.95\columnwidth]{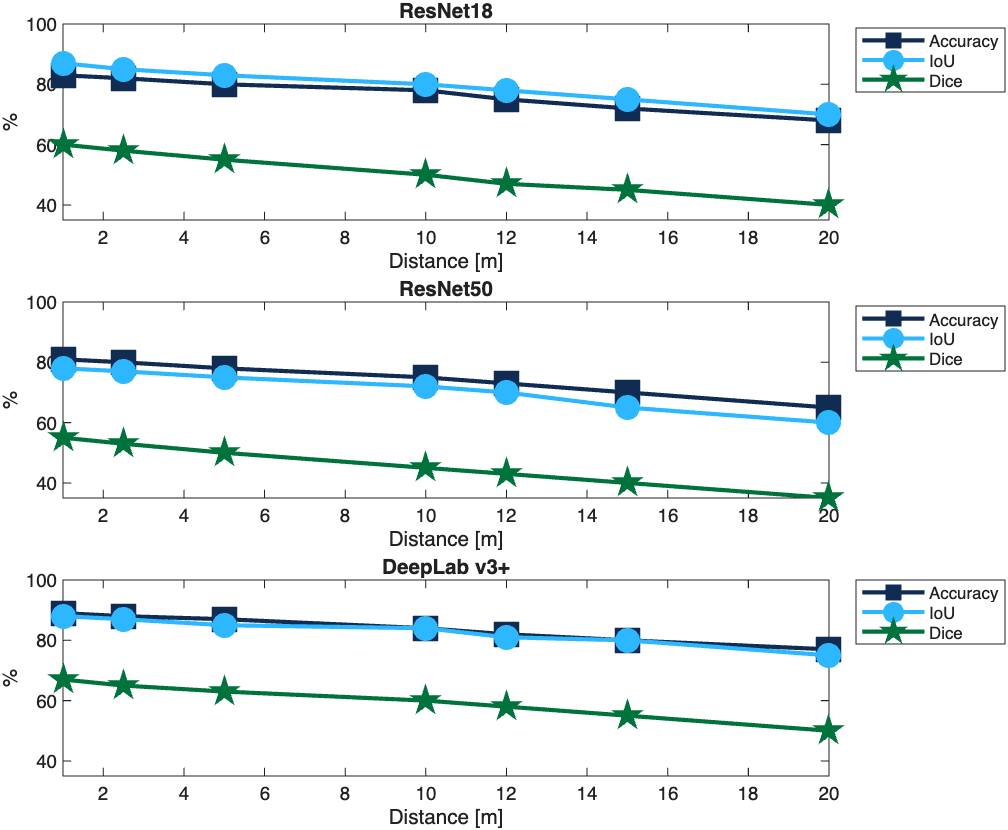}}
\caption{Comparison of SmartBAN classification performance at different transmitter–receiver distances in presence of various other interference signals for various neural network architectures. Accuracy, Dice score, and IoU are shown as functions of distance.}
\label{fig:smartbanMetrics}
\end{figure}

\begin{figure}[!t]
  \centering
  \subfloat[Recognition of all signal types with interference suppression using the ResNet50 network in a synthetic spectrogram. The image on the right illustrates the accurate labelling of all signals present in the spectrogram.]{\includegraphics[width=0.9\columnwidth]{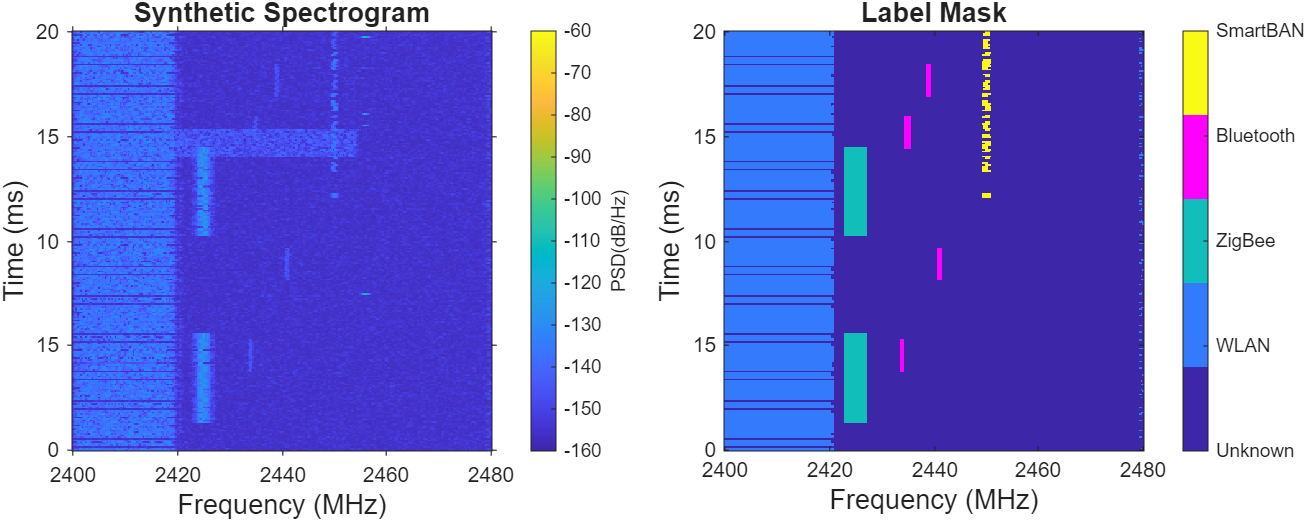}\label{fig:6a}}\\
  \subfloat[
  Accurate detection of signals and boundaries challenging environments with the ResNet50 network. The signals are less clearly defined and exhibit greater distortion.]
  {\includegraphics[width=0.9\columnwidth]{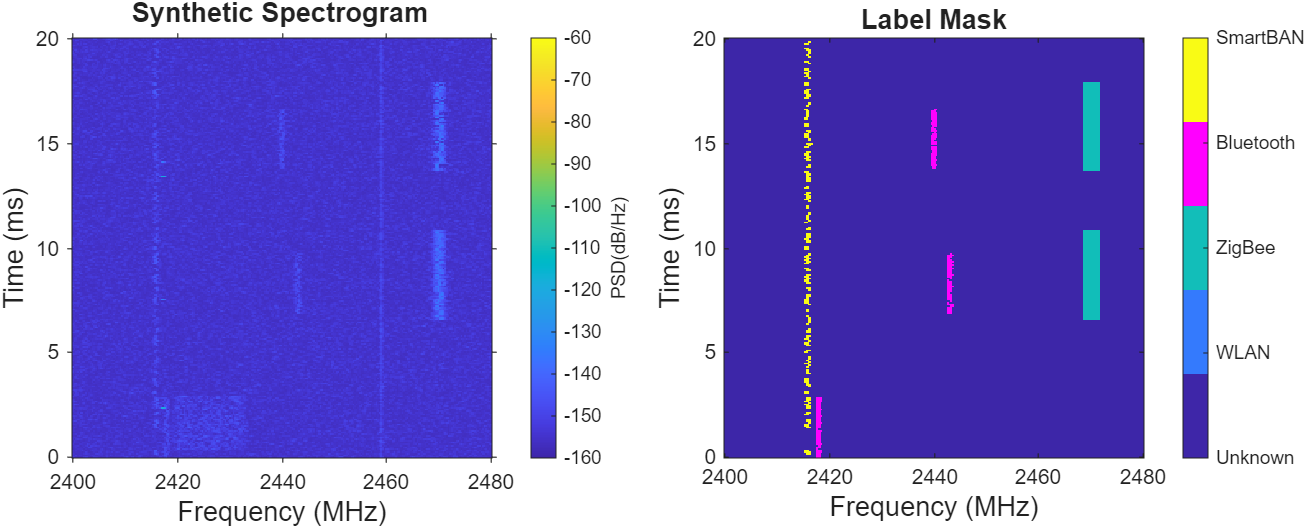}\label{fig:6b}}\\
\caption{ResNet50 network classification performance on synthetic captures with two ISM band conditions.}
  \label{fig:tre_immagini}
\end{figure}


\section{Numerical and Experimental Results}

To validate the proposed framework, network performance was evaluated on both synthetic datasets and real-world RF captures collected from the congested 2.4–2.48 GHz ISM band. The evaluation specifically aimed to assess the framework’s robustness in scenarios where low-power SmartBAN signals are subject to interference from overlapping transmissions.

\subsection{Simulation results}
The performance metrics considered for evaluation are the Accuracy, Intersection over Union (IoU), and Dice Score, defined respectively as
\begin{equation}
\text{Accuracy} = \frac{TP + TN}{TP + TN + FP + FN},
\end{equation}
\begin{equation}
\text{IoU} = \frac{TP}{TP + FP + FN},
\end{equation}
\begin{equation}
\text{Dice} = \frac{2TP}{2TP + FP + FN},
\end{equation}
where \(TP\), \(TN\), \(FP\), and \(FN\) denote the true positives, true negatives, false positives, and false negatives, respectively.


The evaluation on synthetic spectrograms demonstrates consistently high performance across all three networks, as illustrated in Figure~\ref{fig:comparativeAnalysis}. \textit{ResNet-18}, \textit{ResNet-50}, and \textit{DeepLab~v3+} all achieve strong results, showing a balance of classification accuracy and inference time (Table~\ref{tab:prediction_time}). These outcomes confirm the robustness of convolutional architectures for this task.

To gain deeper insight into the network’s behaviour in \textit{SmartBAN} signal classification, a dedicated simulation was conducted. A specialized dataset of spectrograms was generated by placing a reference SmartBAN transmitter at distances of 
\(1,\, 2.5,\, 5,\, 8,\, 10,\, 12,\, 15,\) and \(20\,\mathrm{m}\) from the receiver. 
The dataset was produced following the same procedure described in Section III-B. All transmitters, except for the reference SmartBAN node, were randomly positioned within a circular area of radius \(20\,\mathrm{m}\). 
An indoor environment was assumed, with the receiver located at the origin of the coordinate system. 
The resulting dataset comprises 1000 spectrograms.

The performance in SmartBAN signal classification is summarized in Figure~\ref{fig:smartbanMetrics}, which shows the evolution of network accuracy, Dice score, and F1-score as the transmitter–receiver distance increases. As expected, all models exhibit a gradual degradation in performance due to signal attenuation and increased noise dominance. Nevertheless, \textit{ResNet-50}, \textit{ResNet-18}, and \textit{DeepLab~v3+} show comparable performance across all metrics and distances, indicating similar robustness to channel impairments.

A more detailed analysis on synthetic spectrograms of \textit{ResNet-50} highlights its strong classification capability across all signal types. As shown in Figure~\ref{fig:6a}, the network accurately identifies each signal and delineates their contours with high precision. Furthermore, Figure~\ref{fig:6b} illustrates that the model maintains high reliability even under challenging simulated conditions, where low-power SmartBAN signals are subject to interference from overlapping transmissions. In these environments, interfering signals are effectively disregarded, allowing the true SmartBAN regions to remain clearly identifiable.

\subsection{Experimental Setup and Data Acquisition}

\begin{figure}[!t]
  \centering
  \subfloat[Experimental setup of the over-the-air tests. The wideband receiver (right) monitors the full 80\,MHz ISM band, while the two ADALM-Pluto SDRs (left) transmit a SmartBAN signal and a WLAN signal. Additional Bluetooth traffic is generated by a laptop paired with wireless earbuds.]{\includegraphics[width=0.9\columnwidth]{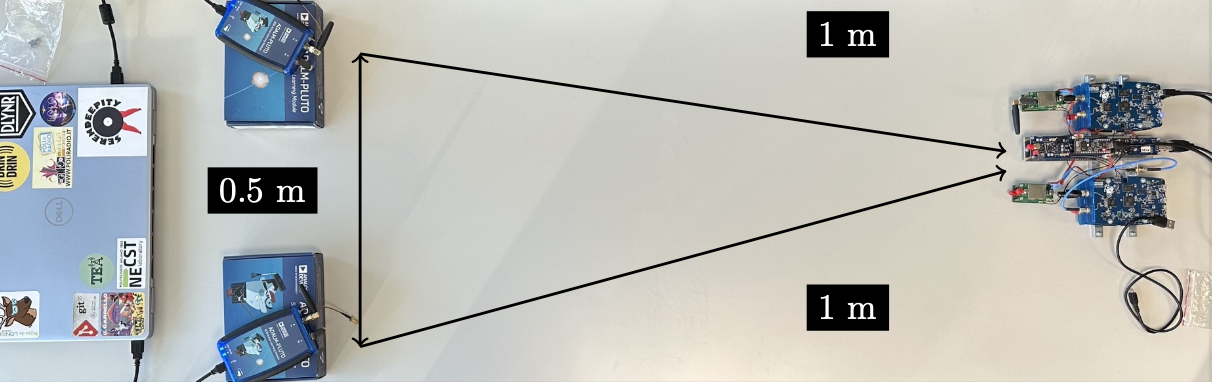}\label{fig:setupExperiment}}
  \\
  \subfloat[
  Example of SmartBAN packet capture and classification.
  The SmartBAN signal is accurately identified and classified, showing clear separation from the interferers.]{\includegraphics[width=0.9\columnwidth]{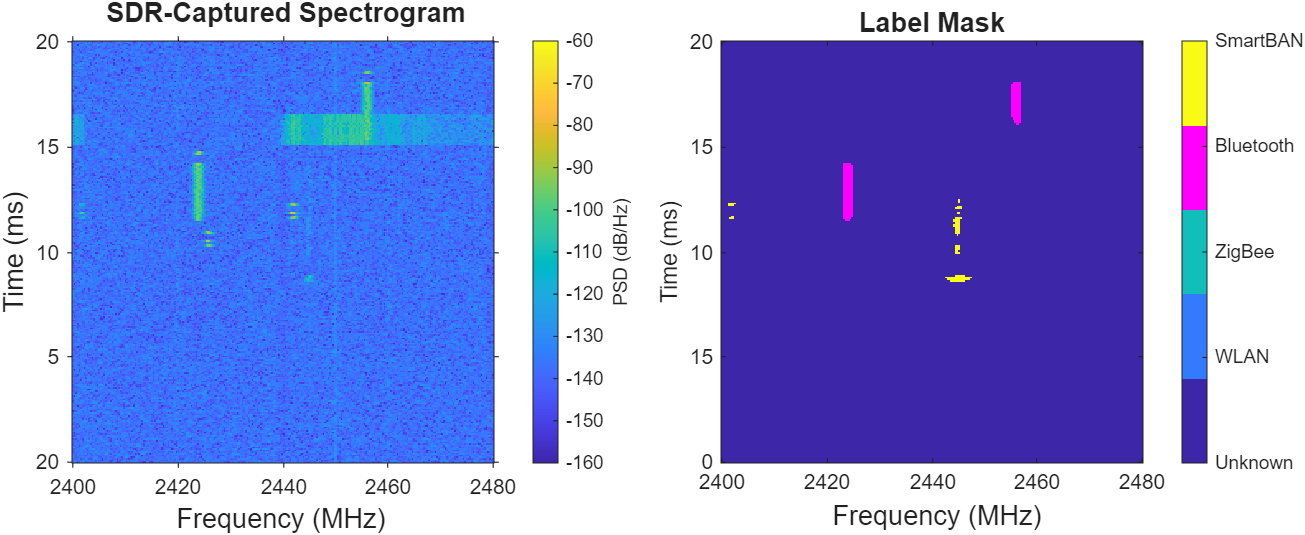}\label{fig:6c}}
\caption{Experimental setup and classification result on a real-world capture at ISM band.}
  \label{fig:tre_immagini}
\end{figure}

Real-world RF signals were captured using Software-Defined Radios (SDRs) configured to acquire multiple adjacent frequency sub-bands covering the 2.4--2.48\,GHz ISM range. To emulate realistic interference conditions, transmissions from WLAN and Bluetooth devices were recorded, together with low-power SmartBAN-like and ZigBee-like bursts.
A dedicated script was developed to randomly generate either a SmartBAN or a ZigBee signal, which was subsequently transmitted using an ADALM-Pluto SDR. In parallel, WLAN and Bluetooth traffic was produced by external consumer devices such as smartphones, tablets, and wireless earphones, creating a realistic coexistence scenario. 

The complete experimental setup is illustrated in Figure~\ref{fig:setupExperiment}. The receiver setup consisted of four synchronized ADALM-Pluto SDRs operating in both time and frequency coherence through dedicated synchronization hardware. To account for hardware bandwidth limitations, the entire ISM band was divided into multiple spectral portions, each assigned to a specific SDR. The individual sub-band captures were then merged to reconstruct a continuous 80\,MHz-wide spectrogram.
All measurements were conducted in a controlled indoor environment populated with at least 1 Bluetooth device, ensuring a realistic and interference-rich context. During each acquisition cycle, the SmartBAN/ZigBee generation script operated continuously, while the receiver array captured the resulting composite RF scene.

For the real-world experiments, the \textit{ResNet-50} architecture was adopted, as it demonstrated superior performance compared to \textit{ResNet-18} and \textit{DeepLab v3+} when applied to real spectrograms. As shown in Figure~\ref{fig:6c}, the network correctly identifies the target SmartBAN beacons and effectively disregards the interferer signals. Notably, the network maintains a high detection accuracy even for low-power SmartBAN signals, which are typically challenging to distinguish as they can be overshadowed by stronger coexisting transmissions. This behavior demonstrates the proposed framework's ability to capture subtle spectral and temporal features, ensuring reliable classification even under unfavorable real-world signal-to-noise conditions.

\section{Conclusion}
This work demonstrates the relevance and potential impact of intelligent RF signal classification for medical Body Area Networks (BANs), particularly in the context of SmartBAN deployments operating in the congested 2.4\,GHz ISM band. The proposed framework successfully integrates synthetic spectrogram generation with real over-the-air acquisitions, enabling both controlled training conditions and realistic evaluation scenarios. On synthetic datasets, the system achieved excellent performance, exceeding 90\% overall accuracy and obtaining excellent results in the classification of the SmartBAN signals across all tested network configurations. When assessed using real SDR-captured signals, the proposed framework maintained consistent, though more moderate, classification performance, reflecting the complexity and variability of practical wireless environments.
These results confirm the feasibility of identifying extremely low-power SmartBAN waveforms in dense spectral conditions, a capability that is essential for interference-aware coexistence management and for ensuring reliable medical data exchange in wearable healthcare systems. Future work will investigate more robust preprocessing and adaptive learning strategies to improve generalization and efficiency in real BAN deployments.

\footnotesize

\bibliography{Bibliografia}

\end{document}